\documentclass[preprint]{aa}
\usepackage{txfonts}
\usepackage{graphicx,subfigure}
\usepackage{epsfig}
\usepackage{epsf}
\usepackage{lscape}
\usepackage{natbib}
\usepackage{rotating}
\usepackage{tabularx}
\usepackage{lscape}
\usepackage{supertabular}

\newcommand{\Healpix}{\mbox{\sc{HEALPix }}}
\newcommand{\Fits}{\mbox{\sc{FITS }}}
\usepackage{url}

\begin{document}
\title{Dark gas in the solar neighborhood from extinction data}

\author{D. Paradis \inst{1,2} \and K. Dobashi \inst{3} \and
  T. Shimoikura \inst{3} \and A. Kawamura \inst{4} \and T. Onishi
  \inst{5} \and Y. Fukui \inst{4} \and J.-P. Bernard \inst{1,2}}
 
\institute{Universit\'e de Toulouse; UPS-OMP; IRAP; Toulouse, France 
\and
CNRS; IRAP; 9 Av. du Colonel Roche, BP 44346, F-31028, Toulouse, cedex
4, France
\and
Department of Astronomy and Earth Sciences, Tokyo Gakugei University,
Koganei, Tokyo 184-8501, Japan
\and
Department of Astrophysics, Nagoya University, Chikusa-ku, Nagoya
464-8602, Japan
\and 
Department of Physical Science, Osaka Prefecture University, Gakuen
1-1, Sakai, Osaka 599-8531, Japan
}
\authorrunning{Paradis et al.}
\titlerunning{Dark gas in the solar neighborhood from extinction data}

\abstract
{When modeling infrared or $\gamma$-ray data as a linear combination
of observed gas tracers,  excess emission has been detected
compared to expectations from known neutral and molecular gas
traced by HI and CO measurements, respectively. This excess might
correspond to additional gas component. This so-called "dark gas"
(DG) has been observed in our Galaxy, as well as the Magellanic
Clouds.}
{For the first time, we investigate the correlation
between visible extinction ($\rm A_V$) data and gas tracers 
on large scales in the solar neighborhood, to detect DG and to verify
our compatibility with previous studies.}
{Our work focuses on both the solar neighborhood ($\rm |b|>10\degr$),
  and the inner and outer Galaxy, as well as on four individual regions:
Taurus, Orion, Cepheus-Polaris, and Aquila-Ophiuchus. Thanks to the
recent production of an all-sky $\rm A_V$ map, we first perform the
correlation between $\rm A_V$ and both HI and CO emission over the most
diffuse regions (with low-to-intermediate gas column densities), to
derive the optimal $\rm\left (  A_V/N_H \right )^{ref}$ ratio. We then iterate the analysis
over the entire regions (including low and high gas column densities)
to estimate the CO-to-H$_2$ conversion factor, as well as the DG mass
fraction.}
{The average extinction to gas column-density ratio in the solar
neighborhood is found to be $\rm \left ( A_V/N_H \right )^{ref}=6.53\times 10^{-22}$
mag. cm$^{2}$, with significant differences between the inner and
outer Galaxy, of about 60$\%$. We derive an average value of the
CO-to-H$_2$ conversion factor of $\rm X_{CO}=1.67\times10^{20}$
$\rm H_2\,cm^{-2}$/(K km s$^{-1}$), with significant variations
between nearby clouds. In the solar neighborhood, the gas mass in the
dark component is found to be 19$\%$ relative to that in the atomic
component and 164$\%$ relative to the one traced by CO. These results
are compatible with a recent analysis of Planck data within
the uncertainties of our measurements. We estimate the ratio
of dark gas to total molecular gas to be 0.62 in the solar
neighborhood. The HI-to-H$_2$ and H$_2$-to-CO transitions appear for
$\rm A_V\simeq0.2$ mag and $\rm A_V\simeq1.5$ mag, respectively, in agreement
with theoretical models of dark-H$_2$ gas.}
{}

\maketitle
\section{Introduction}
The interstellar medium (ISM) is composed of gas and dust, where the
dust component represents about 1$\%$ of the total ISM
mass. Interstellar gas provides the material for star formation,
making it a central part in the life cycle of matter. The gas is
either atomic, molecular, or ionized and these three phases are commonly observed
using the HI 21 cm emission, CO transitions, and either $\rm H_\alpha$ or
free-free emission, respectively. Several studies have reported an
emission excess in the far-infrared (FIR) with respect to the
gas, that is correlated with none of the gas tracers. These excess detections were
first obtained for Galactic regions \citep{Reach94,
Meyerdierks96}. By analyzing the diffuse $\gamma$-ray emission
from the ISM, \citet{Grenier05} concluded that there is an
additional gas phase in our Galaxy called "dark gas'' (DG),
which has a non-negligible mass. This component has also been discovered
in the Magellanic clouds \citep{Leroy07,Bernard08,
RomanDuval10}. \\

The \citet{Bernard11} obtained all-sky maps of the dust
optical depth and, comparing it with the observed gas column density,
constructed a map of the DG distribution covering a large fraction of
the intermediate Galactic latitude sky ($\rm |b|>10\degr$). On average, they
estimated that the mass of DG is 28$\%$ of the atomic mass and 118$\%$ of the molecular gas traced by CO emission in the solar
neighborhood. These results indicate that the DG is detected at
intermediate hydrogen column densities, corresponding to extinctions
values between 0.4 and 2.5 mag. The first possible explanation is that
the DG could be molecular gas, which is not detected using the CO(J=1-0)
transition. It is indeed expected that a layer of pure H$_2$
or dissociated CO should exist around dense clouds
\citep[e.g.][]{Wolfire10} that would be undetectable because at
temperatures below 100 K (which is typical of these environments), the fluxes
from the H$_2$ rotational transitions are too low. However, this layer could also be traced using other
species such as C$^+$ \citep[see][]{Langer10}.
Other possible origins were suggested by the \citet{Bernard11} to
explain the observed departure from linearity between the dust optical
depth ($\tau$) and the observable gas column density, namely: (1) variations
in the dust/gas ratio (D/G), (2) weak CO emission undetected at the
sensitivity of the CO survey, (3) optically thin approximation 
for the HI emission not valid over the entire sky, and (4) formation of
dust aggregates, inducing a higher dust emissivity. For hypothesis (1), the authors concluded that variations in D/G of about
30$\%$ are unlikely in the solar neighborhood.  Assumptions (2) and
(3) were tested by performing the analysis with an upper limit in
the weak CO emission in one case, and with a different HI spin
temperature in the other case. The authors deduced that these two
effects could not account for the whole excess. The last option (4)
can be tested using extinction data, since dust aggregates are
expected to have a higher FIR emissivity than isolated grains, but mostly
unaffected optical properties in the visible and ultra-violet.
Therefore, no substantial $\rm A_V$ excess relative to the gas column
density should be observed in that case. We note that
$\gamma$-ray observations would not detect any excess if the FIR
excess were due to either D/G or dust optical property variations.
\\

The large dust particles that emit in the FIR are also
responsible for the main extinction in the visible and near-infrared
(NIR). Extinction data are therefore well-suited to checking for
the presence of DG. An extinction map covering the whole sky was produced by \citet{Dobashi11} using the Two Micron All Sky Survey Point Source Catalog
(2MASS PSC). This map has since
been significantly improved to correct for the background-star
intrinsic colors \citep{Dobashi11b}.

This study is carried out following the methodology adopted by
\citet{Bernard11}, but using extinction instead of FIR optical
depth. The use of dust optical depth requires us to
determine dust temperature and therefore make assumptions about the
dust emissivity shape and/or mixing effects along the
line-of-sight (LOS). On the other hand, in principle, extinction
directly measures
dust column density. In practice however, large-scale extinction maps derived from star counts have limited accuracy
and suffer from a bias inherent to the stars being intermixed
with the gas, and starlight does not sample the whole LOS.
In this analysis, we use the same gas tracers as in \citet{Bernard11}, so
that the results can be compared directly.

Here, we compare the spatial distribution of the DG seen
in absorption and emission, as well as the DG masses derived from the
two approaches. We also wish to investigate whether the
transition between the HI and H$_2$ phases is consistent with that
found using FIR emission. In addition, we are interested in comparing
our results with theoretical models such as that of \citet{Wolfire10}.

In Section \ref{sec_obs}, we present the dataset used in this analysis,
in Section \ref{sec_corr} we then describe the method we applied to
perform the correlations between extinction data and gas tracers. We
discuss the results in Section \ref{sec_discussion}. Finally, a
summary of our findings is provided in Section \ref{sec_summary}.
\begin{figure*}
\begin{center}
\includegraphics[width=16cm]{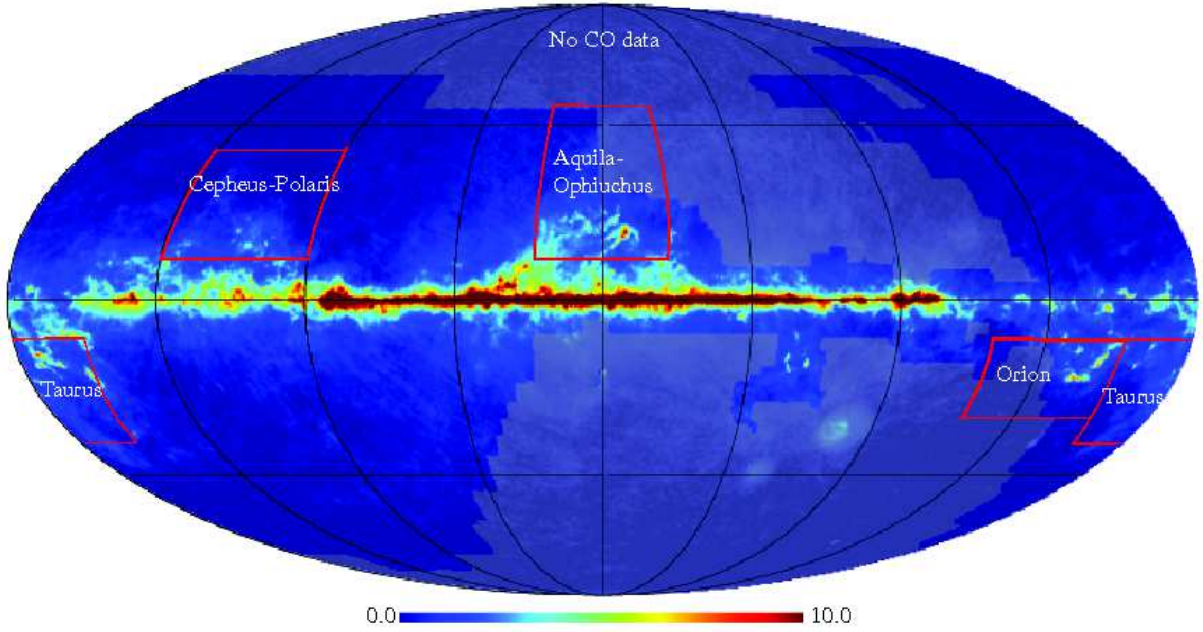}
\caption{Extinction map ($\rm A_V$) in mag. Regions in light blue are
  not covered by the combination of the \citet{Dame01} and
  NANTEN CO surveys, and are not used in our analyses. The red boxes correspond to individual regions
  defined in Section \ref{sec_reg}.\label{fig_avmap}}
\end{center}
\end{figure*}

\section{Observations}
\label{sec_obs}
\subsection{Extinction data} \label{sec_ext_data}
We used color excess maps produced by \citet{Dobashi11} based on the
2MASS PSC \citep{Skrutskie06}.  The maps were derived using a technique
named ``the $X$ percentile method'' \citep{Dobashi08,Dobashi09}, which
is an extension of the well-known NICE method introduced by
\citet{Lada94}.  While the standard NICE method uses the mean color of
stars found in a cell set on the sky to measure the color excess by
dust, the $X$ percentile method utilizes the $X$ percentile reddest
star ($X=100\%$ is the reddest) and is unaffected by the
contamination of unreddened foreground stars.

The combination of the maps of the Galactic plane and both north and south high
Galactic latitudes cover the whole sky with a $1^{\prime}$ grid and a
varying angular resolution (from $1^{\prime}$ to $12^{\prime}$) because of the
"adaptive grid'' mapping technique used \citep{Cambresy99}. This
technique adjusts the resolution to ensure a constant number of stars
in the cells while measuring the color excess, and maintain a flat noise
level all over the sky.

Among the color excess maps presented by \citet{Dobashi11}, we
chose to compare the $\rm E(J-H)$ map measured in the range $\rm 50<X<95$ $\%$
(based on his equation 4) with the gas data, because
  this map is relatively more sensitive than other maps and has
  fewer defects in the regions studied in this paper ($|b|>10^\circ$).
However, we note that the results obtained in the following do not change
significantly when using other maps computed at different $X$
values or from the other color excess available, $\rm E(H-K_{\rm S})$.

Here, we should note that there is a slight but systematic offset in
the original color excess maps of \citet{Dobashi11}; this arises from the
rather ambiguous determination of the background star colors, i.e.,
the mean star colors unreddened by dust, which is needed to determine
the zero point of the color excess.  In general, when deriving a color
excess map on large scales using the NICE method, it is
virtually impossible to determine the background star colors precisely
from the observed stars themselves, because the mean intrinsic star
colors should vary on the sky and there is
no region without dust across the Galactic plane.  For simplicity,
\citet{Dobashi11} assumed a constant mean intrinsic star color in all
directions, as for other all-sky color excess maps derived in a
similar way, e.g., by \citet{Rowles09}; this led to a systematic
offset in the final color excess map.

This varying offset is generally not large, but can cause significant
problems when attempting to detect DG. To reduce these offsets, we use
the new color-excess maps of \citet{Dobashi11b}, who provided a correction to the original color-excess
maps of \citet{Dobashi11} by determining the mean intrinsic star
colors as a function of Galactic coordinates using the
Besan\c{c}on Galactic model \citep{Robin03}, which is one of the
latest Galactic stellar population synthesis model. On the basis of this
model, they generated a star catalog equivalent to the 2 MASS PSC, but
free from any interstellar dust, and calculated how the mean intrinsic
star colors vary across the sky when applying the $X$ percentile
method to the simulated star catalog. They regarded the resulting star
color maps as the background for the color-excess maps derived from
the 2MASS PSC.

In addition, \citet{Dobashi11b} estimated the fraction of total dust
along the LOS that escapes detection by the $X$
percentile method, by simulating the effect of a diffuse dust disk on
the simulated star catalog. The method is known to underestimate the
diffuse dust component extending over a large region along the LOS,
but it can retrieve extinction from individual well-defined dark
clouds more precisely \citep{Dobashi09}.  As a result, they found that
more than 80 $\%$ of the total dust along the LOS should be detected
in the Galactic latitude range $\rm |b|\gtrsim5^\circ$, though 
the effect is more severe in the lower Galactic latitude range (e.g., by $\rm
\sim 50$ $\%$ at $\rm |b| \simeq 0^\circ$).  They derived a map of the
detection rate of the dust disk, which can be used to correct the
color excess maps for this underestimation on large scales.

The color excess maps used in this work were corrected for the
background star colors as well as for the underestimation of the
diffuse and extended dust component.  However, the corrected maps might
still be erroneous close to $\rm |b| \simeq 0^\circ$, especially
around the Galactic center, because of the inhomogeneous detection
limits in the 2MASS PSC and/or unknown stellar populations not taken
into account in the Besan\c{c}on Model. Therefore, when we compare the
maps with the gas data in the following, we restrict ourselves to $\rm
|b|>10^\circ$, where these corrections for the background and the
underestimation are minimized. This also corresponds to the region
studied in \citet{Bernard11}.

Finally, we converted the  $\rm E(J-H)$ map to $\rm A_V$ using
\begin{equation}
A_V=10.9E(J-H),
\end{equation}
where the coefficient corresponds to the empirical reddening law of
\citet{Cardelli89} for $\rm R_V=3.1$. The resulting map (see Figure 1) has an almost
constant noise level of $\rm \Delta A_V \simeq 0.5$ mag.

\subsection{Gas tracers}
The gas tracers used in this analysis are the same as described in
\citet{Bernard11}. Here we provide a summary of the atomic (21 cm
HI emission) and molecular ($^{12}$CO(J=1-0) line) surveys.  We used
the LAB (Leiden/Argentine/Bonn) survey to trace the atomic gas. This
survey is a combination of the Leiden/Dwingeloo survey by
\citet{Hartmann97} with sky observations above -30$\degr$ of Galactic
latitude (at a 36$^{\prime}$ angular resolution), and the IAR
(Instituto Argentino de Radioastronomia) survey
\citep{Arnal00,Bajaja05} of the Southern sky at latitudes below
-25$\degr$ (at a 30$^{\prime}$ angular resolution). The LAB data were
integrated in the velocity range -400$<$$\rm V_{LSR}$$<$400 km
s$^{-1}$.  Assuming that the gas is optically thin, we deduced the
hydrogen column density from the integrated intensity of the HI
emission ($\rm W_{HI}$) using
\begin{equation}
N_{HI}=X_{HI}W_{HI},
\end{equation}
where $\rm X_{HI}$ is the HI integrated intensity to column-density
conversion factor. This factor is taken to be equal to
1.82$\times$10$^{18}$ $\rm H/cm^2/(K\,km\,s^{-1})$ \citep{Spitzer78}.

For the molecular gas, we used the combination of three
$^{12}$CO(J=1-0) line surveys:
\begin{itemize}
\item{The \citet{Dame01} survey, in the Galactic plane, obtained with
both the CfA telescope in the north (at an angular resolution of
8.4$^{\prime}$) and CfA-Chile telescope in the south (at an angular
resolution of 8.8$^{\prime}$). The integrated intensity map was
derived by integrating the velocity range where the CO emission is
significantly detected \citep{Dame11}.}
\item{The unpublished high latitude survey obtained with the CfA
telescope, still observing the northern sky \citep{Dame12}. The data
cube was integrated over 10-20 velocity channels. }
\item{The NANTEN survey obtained from Chile for the intermediate
Galactic latitudes not covered by the \citet{Dame01} survey, at a
2.6$^{\prime}$ angular resolution. This survey is still unpublished,
but a full description of the NANTEN telescope can be found for instance in
\citet{Fukui99}. The total intensity map was
obtained by integrating the data cube over the whole velocity range.}
\end{itemize} 
Each survey was smoothed to a common resolution of 8.8$^{\prime}$
using the convolution by a Gaussian kernel. As quoted in
\citet{Bernard11}, the NANTEN data appear to have 24$\%$ higher values
than those obtained with the CfA telescopes. This discrepancy is still poorly
understood and following \citet{Bernard11}, the NANTEN data were
decreased by 24$\%$, before merging with the CfA survey, to bring them
to the absolute
scale of \citet{Dame01}. The total map used in this analysis is shown in
\citet{Bernard11}, Figure 1. The CO integrated intensity ($\rm
W_{CO}$) to column density conversion factor is derived from the relation
\begin{equation}
N_{H_2}=X_{CO}W_{CO}.
\end{equation}
The value of the CO conversion factor ($\rm X_{CO}$) is still
debated. This factor is also expected to vary over the sky. For this
reason, we made no assumption and derived $\rm X_{CO}$ instead 
from correlations (see Section \ref{sec_corr}).

The contribution of emission in the ionized medium is neglected in this
study. This assumption is reasonable since we do not consider the
Galactic plane emission, which includes most of the ionized medium
emission caused by HII regions. However, its possible
small contribution at high latitudes, which is correlated neither with atomic nor
molecular-traced gas, can be accounted for in the constant term ($\rm
A_V^0$) when performing the correlations (see Equation
\ref{eq_corr}). \\

This work is done at the resolution of the HI data,
i. e. 36$^{\prime}$. All data are projected using the HEALPix
pixelization scheme (Hierarchical Equal Area isoLatitude
Pixelization)\footnote{http://lambda.gsfc.nasa.gov/} with nside=256,
corresponding to a pixel size of 13.7$^{\prime}$. The description of
the projection method used is given in Appendix \ref{sec:drizzeling}.
\begin{table*}[!t]
\begin{center}
\begin{tabularx}{\textwidth}{ l  c  c  c c c c c c c}
\hline
\hline
Region & $\rm \left (\frac{A_V}{N_H}\right )^{ref} $ &  $\rm A_V^0$ &$\rm X_{CO}$  & $\rm M_H^X/M_H^{HI}$&
$\rm M_H^X/M_H^{CO}$ & \multicolumn{2}{c}{$\rm f_{DG}$} \\
 & (10$^{-22}$ mag. cm$^2$) &(10$^{-2}$ mag)&($\rm 10^{20}H_2\,cm^{-2}/(K\,
 km\,s^{-1}$) & & &  This work & From previous works\\
\hline
All $|b|>10\degr$ & 6.53$\pm$ 0.03& 6.16$\pm$0.67 &1.67$\pm$0.08 & 0.19$\pm$0.25 &
1.64$\pm$2.20 & 0.62 & 0.55$^a$ \\
Inner Galaxy & 8.72$\pm$0.04 & 5.81$\pm$0.16 & 2.28$\pm$0.11 & 0.24$\pm$0.38  &
2.43$\pm$3.84 & 0.71 & --\\
Outer Galaxy & 5.46$\pm$0.03& 4.68$\pm$0.75  & 1.67$\pm$0.10 & 0.11$\pm$0.28 &
0.74$\pm$1.92 & 0.43 & --\\
Cepheus-Polaris & 6.01$\pm$0.14 & 7.93$\pm$1.33  & 1.37$\pm$0.34 & 0.04$\pm$0.71 &
0.17$\pm$3.02 & 0.15 & 0.1$^b$ - 0.3$^c$\\
Taurus & 5.33$\pm$0.21& 0.91$\pm$0.01  & 2.27$\pm$0.90 & 0.42$\pm$0.02 & 0.76$\pm$0.03
&  0.43 & 0.3$^b$\\
Orion & 3.41$\pm$0.11 & 7.47$\pm$0.29  & 2.84$\pm$0.92 & 0.45$\pm$0.12 & 0.66$\pm$0.18
& 0.40  & 0.1$^b$\\
Aquila-Ophiuchus & 6.99$\pm$0.18& 11.78$\pm$0.53 & 3.46$\pm$0.91 & 0.59$\pm$0.06 &
1.65$\pm$0.18 & 0.62 & 0.6$^b$\\
\hline
\end{tabularx}
\caption{Derived parameters and their 1-$\sigma$ uncertainties, computed
over different regions, for $\rm |b|>10\degr$.\label{tab_param}}
\end{center}
$^a$ From \citet{Bernard11} \\
$^b$ From \citet{Grenier05}\\
$^c$ From \citet{Abdo10}\\
\end{table*}

\section{Extinction/gas correlation}
\subsection{$\rm \left ( A_V/N_H \right )^{ref}$ and $\rm X_{CO}$ determination}
\label{sec_corr}
In \citet{Bernard11}, the reference value of the ratio of
the dust optical depth to the gas column densities ($\rm \left (
  \frac{\tau}{N_H}\right )^{ref} $) was determined from the correlation between $\tau$ and the gas
tracers at low gas-column density. We followed the same type of analysis,
and modeled the extinction $(A_V^{mod})$ in the portion of the sky covered by the
infrared, atomic, and CO-traced surveys, for $\rm |b|>$10$\degr$,
as
\begin{equation}
\label{eq_corr}
A_V^{mod}=\left ( \frac{A_V}{N_H} \right ) ^{ref} \left ( N_H^{HI} +2X_{CO}W_{CO} \right )+A_V^0,
\end{equation}
 where $\left ( \frac{A_V}{N_H} \right ) ^{ref} $ is the reference value of
the ratio of visible extinction to gas column density,
and $ A_V^0$ is a constant, which can account for ionized gas and/or the offset in the extinction map.

However, the $\rm A_V$ map is particularly noisy at low column
density, which makes the estimate of $\rm \left (
  \frac{A_V}{N_H}\right )^{ref} $ difficult. We therefore adopted a
slightly different approach from \citet{Bernard11}. Our
analysis was done in two steps.
We first selected regions with a limited amount of DG by
identifying regions in Figure 8 of \citet{Bernard11} with DG
column densities $\rm N_H^X$ such that $\rm N_H^X/N_H<0.05$ and we
also imposed a low CO content ($\rm W_{CO}<0.2$ K km s$^{-1}$).  
For this region, which we refer to as ``no DG'', we derived the best-fit
parameters $\rm \left (\frac{A_V}{N_H} \right )^{ref}$ and $\rm
A_V^0$ using Eq.\,\ref{eq_corr}, but did not attempt to derive $\rm
X_{CO}$ in this first step because the considered regions have little CO
emission.
Second, we repeated the same analysis including all pixels of the
considered region included in the gas surveys (we refered to this
region as ``all''), while imposing the $\rm \left (\frac{A_V}{N_H}
\right )^{ref}$ and $\rm A_V^0$ values derived in the first step and determined the best-fit value for $\rm X_{CO}$. We also computed the mass of DG from the
difference between the best correlation and the data at that stage.
The procedure above was applied to both the whole high Galactic latitude sky
(see Sec.\,\ref{sec:highlat}) and a set of individual regions (see
Sec.\,\ref{sec_reg}), and in all cases was restricted to $\rm
|b|>$10$\degr$.

\begin{figure*}[!t]
\begin{center}
\includegraphics[width=16cm]{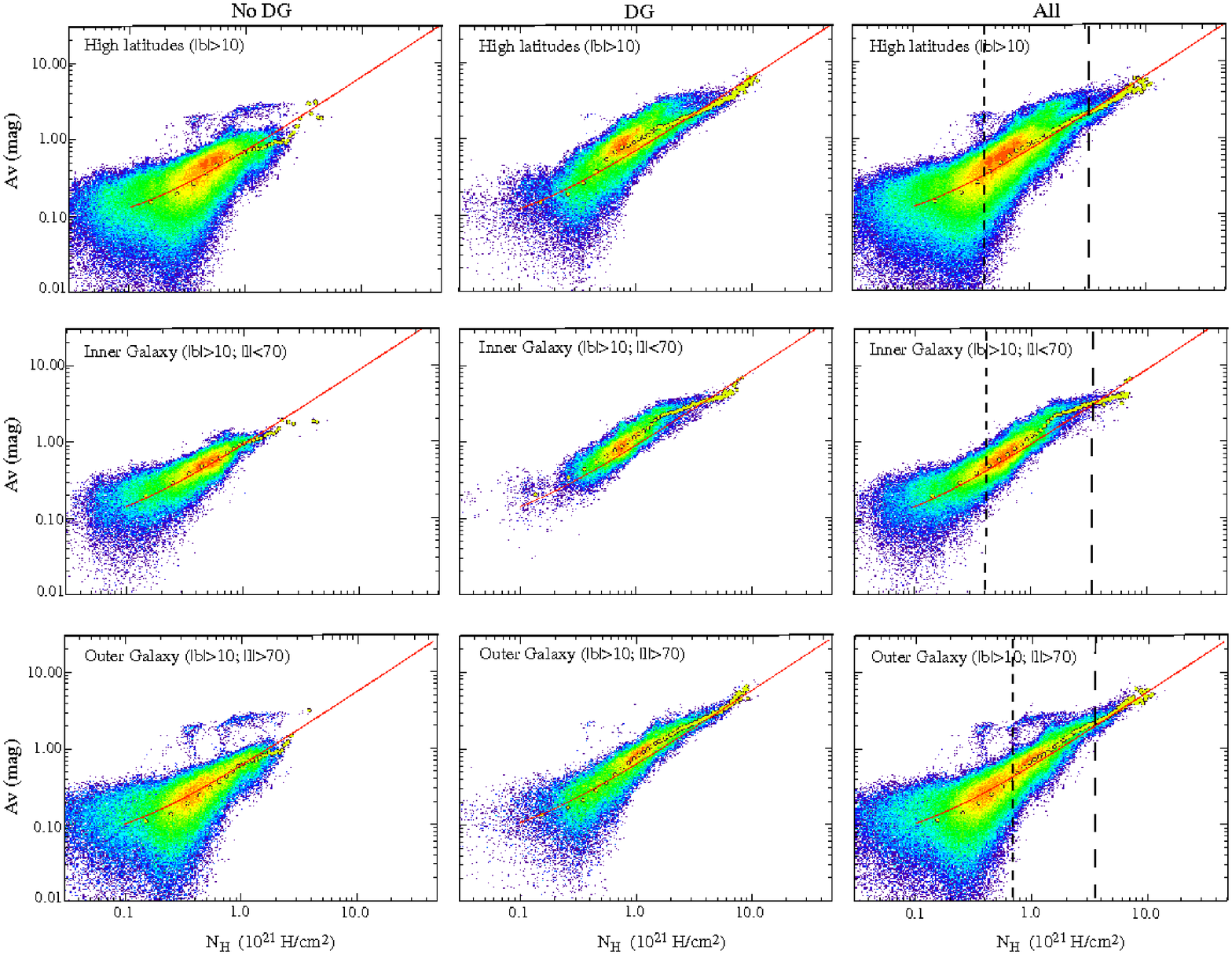}
\caption{Correlation plots between extinction data ($\rm A_V$) and the
total gas-column density ($\rm N_H=N_H^{HI}+2X_{CO}W_{CO}$) excluding the Galactic plane
($\rm |b|>10\degr$): high latitude sky (top panels), inner Galaxy
(middle panels) and outer Galaxy (bottom panels). Left, middle, and
right panels correspond respectively, to regions where no DG is
detected (``no DG'') according to \citet{Bernard11}, where strong DG is detected (``DG''), and
the entire region (``All'') (see Sec.\,\ref{sec_corr}). The color scale
represents the density of pixels on a log scale. The yellow dots show
the correlation binned in $\rm N_H$. The red lines show the best-fit
linear correlation, constrained for the whole data in the ``no DG'' region. The short-dashed and long-dashed lines indicate
the position of a departure from a linear fit, and the position where
the correlation recovers a linear behavior,
respectively.\label{fig_corr_all}}
\end{center}
\end{figure*}

\begin{figure*}[!t]
\begin{center}
\includegraphics[width=16cm]{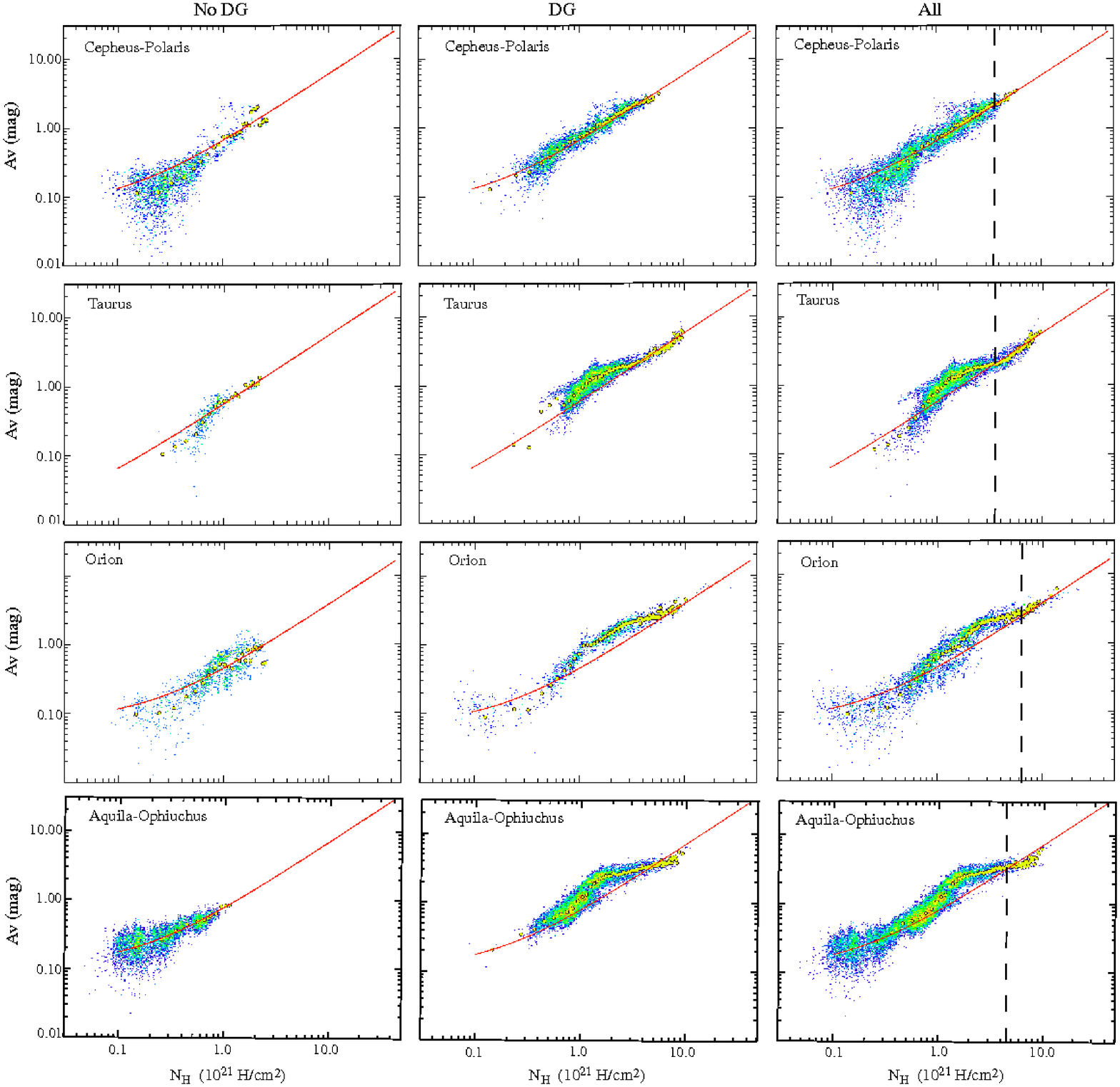}
\caption{Correlation plots between extinction data ($\rm A_V$) and the
total gas column density ($\rm N_H=N_H^{HI}+2X_{CO}W_{CO}$) in various regions excluding the
Galactic plane ($\rm |b|>10\degr$), from top to bottom the
Cepheus-Polaris, Taurus, Orion, and Aquila-Ophiuchus regions. Left,
middle, and right panels correspond respectively, to regions where no DG is
detected (``no DG'') according to \citet{Bernard11}, where strong DG is detected (``DG''), and
the entire region (``All'') (see Sec.\,\ref{sec_corr}). The color
scale represents the density of pixels on a log scale. The description
of the yellow dots, red, and dashed lines is given in the caption of
Figure \ref{fig_corr_all}. \label{fig_corr_regions}}
\end{center}
\end{figure*}

Results of the correlations are presented in Table
\ref{tab_param}. The best-fit parameters were obtained from $\chi^2$
minimization. Parameter uncertainties were derived from the difference
between the minimum and maximum values of the parameters contained in
interval $\Delta \chi^2$, corresponding to a confidence level of
68$\%$. We note that, for the purpose of illustrating the $\rm A_V$-gas
correlations (see Fig.\,\ref{fig_corr_all} and
\ref{fig_corr_regions}), we also carried the correlation in a region
with $\rm N_H^X/N_H>0.05$, which we refered to as the ``DG'' region.

\subsubsection{High Galactic latitude regions ($\rm |b|>10\degr$)}
\label{sec:highlat}
The correlation was first established across the whole high Galactic latitude sky
($\rm |b|>10\degr$) characterizing the solar neighborhood.
Results are displayed in Figure \ref{fig_corr_all} (top panels).
We note that the linear fits (slope=$\rm \left (\frac{A_V}{N_H} \right
)^{ref}$) shown in the figures (red lines) were constrained using the
whole data in the ``no DG'' region, and not the binned version of the
data shown by yellow dots. The corresponding lines are curved
  as plotted on a log-log scale, because of the constant term.
The investigation of this ``no DG'' region leads to an $\rm \left (
  A_V/N_H \right )^{ref}$ ratio of
6.53$\times$10$^{-22}$ mag.cm$^{2}$. This value is close to the
reference value of 5.34$\times$10$^{-22}$ obtained by
\citet{Bohlin78}, which is relevant to the solar neighborhood and the
Galactic plane. We found that the optimal
$\rm X_{CO}$ is close to 1.67 $\times$10$^{20}$ $\rm
H_2\,cm^{-2}/(K\, km\,s^{-1})$. This value is lower than the findings
of \citet{Bernard11}, who derived an averaged $\rm
X_{CO}$=2.54$\times$10$^{20}$ $\rm H_2\,cm^{-2}/(K\, km\,s ^{-1})$, but
in better agreement with the Galactic average of 1.9 $\times$10$^{20}$
$\rm H_2\,cm^{-2}/(K\, km\,s ^{-1})$ \citep{Strong96} and the
value of 1.8 $\times$10$^{20}$ $\rm H_2\,cm^{-2}/(K\, km\,s ^{-1})$
derived by \citet{Dame01} for $\rm |b|>5\degr$.

Looking at Figure \ref{fig_corr_all}, one can discern a different
behavior in the $\rm A_V$ versus $\rm N_H$ plots (with $\rm N_H=N_H^{HI}+2X_{CO}W_{CO}$), when comparing
the ``DG'' and ``no DG'' regions at intermediate and high $\rm A_V$. In
the first case, looking at the binned data points the correlation shows an excess between $\rm N_H \simeq
4\times 10^{20}$ cm$^{-2}$ ($\rm A_V$=0.2 mag) and $\rm N_H \simeq
3\times 10^{21}$ cm$^{-2}$ ($\rm A_V$=1.5 mag) compared to a linear
fit, whereas we do not in the second case. We note that, owing to the low
signal-to-noise ratio of the extinction data at low column densities, the location of these transitions
were determined by eye here, while they were inferred from a fit in
\citet{Bernard11}.
These results imply that there is an additional gas phase, with a
spatial distribution that is similar to that determined using the
Planck data \citep{Bernard11}. For comparison, these authors estimated
that the excess appears at $\rm A_V$ ranging from 0.4 to 2.5
mag. However, results obtained by \citet{mamd11}, by analyzing the
diffuse ISM and Galactic halo, revealed some residual emission in
the infrared/HI correlation, between $\rm N_H\simeq 3 \times 10^{20}$
cm$^{-2}$ ($\rm A_V\simeq$0.15 mag) and $\rm N_H\simeq 4 \times
10^{21}$ cm$^{-2}$ ($\rm A_V\simeq$2.0 mag). Our results are in-between these two findings based on Planck data.

The column density of the dark component ($\rm N_H^X$) can be
computed as the difference between the observed and
the best-fit ($\rm A_V^{mod}$) extinction per unit column density,
over the entire region
\begin{equation}
N_H^X=\frac{A_V-A_V^{mod}}{(A_V/N_H)^{ref}}.
\end{equation} 
The total mass of hydrogen within the DG
component ($\rm M_H^{X}$), the atomic ($\rm M_H^{HI}$), and CO-traced
molecular gas ($\rm M_H^{CO}$), is estimated in the same way as in
\citet{Bernard11} 
\begin{equation}
M_H^{X,HI,CO}=m_H D^2\Omega_{pix}\sum N_H^{X,HI,CO},
\end{equation}
where $\rm D$ is the distance to the gas, $\rm m_H$
the hydrogen atom mass, and $\rm \Omega_{pix}$ the pixel solid
angle. As in \citet{Bernard11}, we assumed the same distance to all gas components.
We obtained $\rm M_H^X/M_H^{HI}$=0.19$\pm$0.25 and
$\rm M_H^X/M_H^{CO}$=1.64$\pm$2.20. These values are expected to be
both overestimated and underestimated, respectively, since the scale-height
of the HI component is larger than that of the molecular one, with a probable
intermediate scale-height for the DG component. The formal uncertainty in the dark gas mass
is large and essentially results from the assumed uncertainty in $\rm
\Delta A_V/A_V$ and from our assumption that this uncertainty is
an absolute error and therefore does not scale down with the number
of independent measurements. This leads to a large uncertainty in the
DG mass when most pixels in a region have low $\rm A_V$ values, but it is
likely that our assumption about the nature of the errors
overestimates the uncertainty.

 Our DG mass is lower than found in \citet{Bernard11}, who deduced
$\rm M_H^{X}/M_H^{HI}\simeq0.28\pm0.03$ and $\rm
M_H^{X}/M_H^{CO}\simeq1.18\pm 0.01$, but consistent with it, taking
our large uncertainties into account.  With the use of their $\rm
X_{CO}$ value, we would obtain $\rm M_H^{X}/M_H^{CO}\simeq0.73$. Our
results lead to a DG mass equal to 17$\%$ of the total observed mass,
close to the value of 22$\%$ derived by \citet{Bernard11}. Our study
using extinction data therefore reveals slightly less DG mass than the same analysis based on FIR data. The difference between the two
analyses is however insignificant owing to large uncertainties for
this region of the sky.

\subsubsection{Inner/outer Galaxy}

We carried out the same analysis for the inner ($\rm |l|<70\degr$)
and outer ($\rm |l|>70\degr$) Galaxy, at high Galactic latitudes ($\rm
|b|>10\degr$). A difference appears in the $\rm \left ( A_V/N_H \right
)^{ref}$ ratio, with a higher value being found in the inner Galaxy ($\rm \left (
  \frac{A_V}{N_H}\right )^{ref} =8.72\times 10^{-22}$ mag.cm$^{-2}$) than the outer
Galaxy ($\rm  \left ( \frac{A_V}{N_H}\right )^{ref} =5.46\times 10^{-22}$ mag.cm$^{-2}$).  This result
could illustrate the true large-scale variations in the dust abundance
and reflect the large-scale metallicity gradient in our Galaxy.

From Figure \ref{fig_corr_all}, one can see that the excess is more
prominent in the inner Galaxy than the outer Galaxy. In addition, we note
that the outer Galaxy appears to be more noisy at low gas column
densities. This is likely because the outer Galaxy
  hosts fewer stars than the inner Galaxy, inducing larger
  uncertainties in the resulting extinction map. The reconciliation between the data and the linear fit,
corresponding to the H$_2$-to-CO transition in the case of
molecular DG, seems to occur at $\rm N_H\simeq 3-3.5\times 10^{21}$ cm$^{-2}$
($\rm A_V\simeq$1.5-1.75 mag) in both the inner and the outer Galaxy.

In term of the DG mass, each part of the Galaxy has a different
behavior. The inner regions contain a more enhanced DG component, with ratios $\rm M_H^{X}/M_H^{HI}$ and $\rm
M_H^{X}/M_H^{CO}$ of about 0.24 and 2.43, respectively. In contrast,
the DG mass in the outer Galaxy represents only 11$\%$ of the
gas in the atomic phase and 74$\%$ of that in the CO-traced
phase. This result indicates that the additional gas phase is
essentially concentrated in the inner Galaxy. In a similar way, the
$\rm X_{CO}$ factor appears to be larger in the inner Galaxy ($\rm
X_{CO}=2.28\times 10^{20}$ $\rm H_2\,cm^{-2}/(K\, km\,s ^{-1})$) than
its outer parts ($\rm X_{CO}=1.67\times 10^{20}$ $\rm
H_2\,cm^{-2}/(K\, km\,s ^{-1})$).

\subsubsection{Individual regions}
\label{sec_reg}
Performing the analysis on individual regions requires the
availability of enough pixels with and without DG, taking into account
the limited extent of the CO survey. We were able to explore the $\rm
A_V$-$\rm N_H$ correlation in the following four regions:
\begin{itemize}
\item{Taurus ($200\degr<l<160\degr$; $-35\degr<b<-10\degr$);}
\item{Orion ($240\degr<l<200\degr$; $-30\degr<b<-10\degr$);}
\item{Cepheus-Polaris ($135\degr<l<90\degr$; $10\degr<b<40\degr$);}
\item{Aquila-Ophiuchus ($|l|<20\degr$; $10\degr<b<48\degr$);} 
\end{itemize}
The results of the correlations for individual regions are presented in
Figure \ref{fig_corr_regions}. Each environment has different dust properties,
with $\rm  \left ( \frac{A_V}{N_H}\right )^{ref}$ ranging between 3.41$\times 10^{-22}$ and 6.99
$\times 10^{-22}$ mag cm$^2$. Moreover, these regions are of great
interest because they have distinctive gas properties. The $\rm X_{CO}$
factor in Orion and Aquila-Ophiuchus are larger than
in the other regions. In addition, the former regions have a
substantially large DG mass ratio relative to the atomic hydrogen, with ratios of 0.45 and 0.59,
respectively. These large amounts of DG in Orion and Aquila-Ophiuchus
are visible in Figure \ref{fig_corr_regions} as a
clear, large excess in the $\rm A_V$-$\rm N_H$
plot. We checked that the excess was not due to
  contamination by young stellar objects (YSOs). The color
  excess maps may indeed overestimate the true $\rm A_V$ if there is
  significant contamination by young stellar objects (YSOs). In these
two regions, we therefore also performed the correlation between extinction
maps in the J band ($\rm A_J$) and the gas tracers, because YSOs
would induce an underestimation in the extinction maps
(as opposed to the color excess maps). These results indicate that the large amount
of DG is not caused by the presence of YSOs.  
\citet{Grenier05} obtained a ratio $\rm M_H^X/M_H^{HI}$ that is even
larger than our results, in the Aquila-Ophiuchus-Libra region, with a ratio of 1. In
contrast, they derived a value of 0.14 in Orion. They found a
ratio of 0.33 for the Taurus-Perseus-Triangulum region, close to
our value of 0.42 in the Taurus region. We note however that the
individual regions defined in \citet{Grenier05} are not exactly the
same as the ones used in this paper. The $\rm A_V$-$\rm N_H$ plot for
Cepheus-Polaris does not display any noticeable excess.  When
comparing the DG mass with the CO one in the different
clouds, we observe significant variations. We note values ranging from
 0.17 in the Cepheus-Polaris region to 1.65 in Aquila-Ophiuchus, with
values of 0.66 and 0.76 in Orion and Taurus. These ratios are
however highly dependent on the value of the $\rm X_{CO}$ factor. The
low uncertainty in the dark mass of the Taurus region is caused by
the relatively large $\rm A_V$ values. The absolute uncertainty
$\rm \Delta A_V/A_V$ is larger in Cepheus-Polaris which corresponds to a
large number of pixels at low $\rm A_V$, inducing a considerable mass
uncertainty.

The H$_2$-to-CO transition, which we determined by eye, appears at $\rm
N_H\simeq 3.5\times 10^{21}$ cm$^{-2}$ ($\rm A_V \simeq 1.75$ mag) in
Taurus, $\rm N_H\simeq 6\times 10^{21}$ cm$^{-2}$ ($\rm A_V \simeq
3.0$ mag) in Orion, $\rm N_H\simeq 3.5\times 10^{21}$
cm$^{-2}$ ($\rm A_V \simeq 1.75$ mag) in Cepheus-Polaris, and $\rm N_H\simeq 4\times 10^{21}$ cm$^{-2}$ ($\rm A_V \simeq 2.0$ mag)
in Aquila-Ophiuchus. This transition seems to vary with
environment. The position of the HI-to-H$_2$ transition is uncertain
because of the noise in the extinction data at low gas-column
densities. For this reason, we do not discuss its value further in this
paper.

\section{Discussion}
\label{sec_discussion}
The two quantities $\rm \tau$ and $\rm A_V$ are in principle
proportional. However, the estimate of the dust optical depth $\rm
\tau$ requires knowledge of the dust equilibrium temperature, requiring an
additional step in the study of the dust/gas ratio, and
additional assumptions that could bias the results. For instance, a
single dust temperature is often assumed along the LOS. In addition,
the temperature estimate depends on the assumption made about the
emissivity spectral index $\beta$. In that sense, the use of
extinction data is more straightforward. However, extinction data
suffer from several defects, especially when producing the color
excess map. Even if the $X$ percentile method used to generate
the map is a promising method, the resulting map can still be biased
by missing extinction, inducing variations in the detection
rate. The detection rate, i.e., the fraction of the detected A$\rm _V$ to the true total A$\rm
  _V$ integrated along the LOS, could vary as a function of both the
  Galactic coordinates and the distance to the clouds. To
  determine how much extinction we should miss in the adopted A$\rm _V$
  map, we have carried out a test by applying the $X$ percentile
  method to some artificial clouds set in the distribution of stars
  generated using the Besan\c{c}on Model \citep{Robin03}. We first
  prepared a spherical cloud having a Gaussian density distribution
  with a size of 6 pc (at FWHM) and a total extinction along
  the LOS of $\rm A_V=10$ mag at the center of the cloud. We then located the
  cloud at realistic coordinates and distances of some nearby clouds
  in the simulated star distribution , e.g., at (l,b)=(174.0$\degr$, -13.5$\degr$)
  and at D=150 pc for the Taurus cloud, and performed the $X$
  percentile method exactly in the same way as \citet{Dobashi11} did
  using the 2MASS PSC. Our results indicate that most of the individual
  clouds analyzed here are well-detected with a detection rate larger
  than 90$\%$ (i.e., the model cloud with $\rm A_V=10$ mag is detected as
  $\rm A_V>9$ mag), except for Orion whose detection rate is
  $\simeq$76$\%$. General results show that nearby
clouds (D$<$500 pc), which represent most of the clouds at $\rm
|b|>10\degr$ in the $\rm A_V$ map, are clearly detected with a detection
rate $>80\%$ for all directions. The low $\rm \left ( A_V/N_H \right )^{ref}$ ratio derived
in Orion could be the result of underestimating $\rm A_V$ in this
region. Apart from this region, the $\rm \left ( A_V/N_H \right )^{ref}$ ratios in Table
\ref{tab_param} appear equal or larger than the reference value given
by \citet{Bohlin78}, confirming that the $X$ percentile method does
not miss a substantial amount of extinction.

The determination of the $\rm X_{CO}$ conversion factor differs
from one analysis to another. Our average value of 1.67$\times
10^{20}$ $\rm H_2\,cm^{-2}/(K\, km\,s ^{-1})$ is close to the Galactic
average \citep{Strong96} and the value derived by \citet{Dame01} for
$\rm |b|>5\degr$. A decrease in $\rm X_{CO}$ is observed from the
inner to the outer high Galactic latitude sky. Most of the clouds observed in
the Galactic extinction map at $\rm |b|>10\degr$ are nearby local clouds, at a
distance of about 200-500 pc, so we do not attribute the variations to
the metallicity gradient \citep{Rolleston00}, but instead to local variations. \citet{Arimoto96}, using CO data and virial
masses, derived an average value near Sun of 2.8$\times10^{20}$ $\rm
H_2\,cm^{-2}/(K\,km\,s ^{-1})$, with variations ranging from
2.09$\times10^{20}$ to 3.74$\times10^{20}$ $\rm
H_2\,cm^{-2}/(K\,km\,s ^{-1})$ with increasing Galactocentric
radius. \citet{Abdo10} discovered a similar behavior, but with
significant variations ranging from 0.87$\times10^{20}$ $\rm
H_2\,cm^{-2}/(K\,km\,s ^{-1})$ in the Gould Belt to 1.9$\times10^{20}$
$\rm H_2\,cm^{-2}/(K\,km\,s ^{-1})$ in the Perseus arm. Finally,
the \citet{Bernard11} estimated an average value of 2.54$\times10^{20}$
$\rm H_2\,cm^{-2}/(K\,km\,s ^{-1})$ in the solar neighborhood. We also
note that some of the studies included the dark component in their
derivation of $\rm X_{CO}$.

In their model to estimate the DG mass, \citet{Wolfire10} predicted an
HI-to-H$_2$ transition located at $\rm A_V\simeq 0.2$ mag, which
agrees with our findings. This result is however slightly lower
than the value derived using FIR data ($\rm A_V\simeq 0.4$ mag)
following \citet{Bernard11}. However, in the latter case the
HI-to-H$_2$ transition was determined by $\chi^2$ minimization,
whereas in our study it was done by eye, which can induce large
uncertainties.

Moreover, the model predictions suggest a constant fraction of the
molecular mass in the dark component of about 0.3 for an average
extinction $\rm A_V$ around 8 mag. In cases of decreasing extinction,
they found an increase in the dark mass fraction. In the framework of
their model, this would indicate that a larger fraction of molecular
gas is located outside the CO region. This fraction
($\rm f_{DG}$) is computed as
\begin{equation}
f_{DG}=\frac{M_H^X}{M_H^X+M_H^{CO}}.
\end{equation}
Values of $\rm f_{DG}$ for each region are provided in Table
\ref{tab_param}. Over the entire high Galactic latitude sky, we derived $\rm
f_{DG}=0.62$, with a larger value in the inner Galaxy ($\rm
f_{DG}=0.71$) with respect to the outer Galaxy ($\rm
f_{DG}=0.43$). For comparison, using the same definition of this
quantity, the \citet{Bernard11} obtained a value of $\rm f_{DG}\simeq$0.55
for the solar neighborhood. The Taurus and Aquila-Ophiuchus regions
have a DG fraction of 0.43 and 0.62, respectively. These values
are in good agreement with the results obtained by \citet{Grenier05}
(see Table \ref{tab_param}) in equivalent regions, with fractions of
0.3 and 0.6, for the Taurus and Aquila-Ophiuchus-Libra regions. These
authors derived $\rm f_{DG}$ as low as 0.1 for
Cepheus-Cassiopeia-Polaris and Orion, whereas \citet{Abdo10} deduced a
value of 0.30 for Cepheus. We obtained $\rm f_{DG}$ equal to 0.15 and
0.40 in Cepheus-Polaris and Orion.

All these studies illustrate the existence of variations in the
dark-gas mass fraction. In our study, we have excluded the Galactic plane,
where the most massive and bright molecular clouds are located. At
  the 36$^{\prime}$ angular resolution, most of the pixels considered in our analysis have
extinctions lower than 10 mag. In this case, $\rm f_{DG}$ values larger
than 0.3 are expected following \citet{Wolfire10}, and this is indeed
what is observed here.

Dust in the form of aggregates is likely to be more emissive than isolated grains, but aggregation is not expected to substantially
affect absorption properties in the visible \citep{Kohler11} .
The detection of significant departure from linearity between
extinction data and gas column densities, especially in Orion or
Aquila-Ophiuchus, which exhibit the largest excess in Figure
\ref{fig_corr_regions}, indicates that dust emissivity changes induced
by grain aggregation is an unlikely explanation of the observed excess.
As a consequence, grain coagulation can only be responsible for a small
amount of the observed excess, and may explain some of the differences in the
studies based on FIR data \citep{Bernard11} and visible extinction
data (this work). Moreover, our results agree with the model
prediction of \citet{Wolfire10}, favoring the hypothesis of pure H$_2$
gas (without CO molecules), surrounding the CO regions.

\section{Summary}
\label{sec_summary}
Using a recently revised all-sky extinction map, we have examined
the correlation between the extinction and gas column densities derived
from HI and CO observations. Our results can be summarized as follows:
\begin{itemize}
\item{We have measured an excess of extinction at intermediate column
densities, relative to a linear-fit between $\rm A_V$ and observable
gas. This excess is observed over the whole high Galactic latitude sky, with a
predominant amount being found toward the inner regions ($\rm |l|<70\degr$), and
in particular in Aquila-Ophiuchus. This result confirms the
recent detection of DG in the Planck data and implies that the
effect in the FIR is not due to changes in the dust optical properties
caused by dust aggregation.}
\item{We have derived an average dust extinction to gas ratio of $\rm
\left ( A_V/N_H \right )^{ref}=6.53\times10^{-22}$ mag cm$^{-2}$ in the solar neighborhood,
for an average $\rm X_{CO}=1.67\times10^{20}$ H$_2$cm$^{-2}$/(K km s$^{-1}$). The inner/outer Galaxy exhibits a higher/lower $\rm
\left ( A_V/N_H \right )^{ref}$ ratio (8.72$\times10^{-22}$/5.46$\times10^{-22}$ mag
cm$^{-2}$), associated with a higher/lower $\rm X_{CO}$ factor
(2.28$\times10^{20}$/1.67$\times10^{20}$ H$_2$cm$^{-2}$/(K km s$^{-1}$). A significantly larger value, more than twice the value
derived in the solar neighborhood is observed in the Aquila-Ophiuchus
region. In contrast, the lowest $\rm X_{CO}$ value is found in
Cepheus-Polaris.}
\item{The results of our analysis agree with the theoretical
predictions of modeling of the dark H$_2$ gas obtained to help explain
the observed departure from linear-fits in the correlations. }
\item{The DG mass derived is 19$\%$ of the atomic mass and 164$\%$ of that
in the CO-traced gas in the solar neighborhood. It can reach up to
59$\%$ of the atomic gas in the Aquila-Ophiuchus region. Our DG mass
estimates are slightly lower than that derived from FIR Planck data, but the
difference is insignificant. }
\item{We have estimated the fraction of the molecular mass in the dark
component ($\rm f_{DG}$) and found an average value of 0.62 in the
solar neighborhood, with variations going from 0.43 to 0.71 in the
outer and inner Galaxy. We have derived $\rm f_{DG}$=0.15, 0.43, 0.40,
and 0.62 in the Cepheus-Polaris, Taurus, Orion, and Aquila-Ophiuchus
regions, respectively.}
\item{The HI-to-H$_2$ and H$_2$-to-CO transitions appear for $\rm
A_V\simeq0.2$ and $\rm A_V\simeq1.5$ mag in the solar neighborhood,
with variations from regions to regions. }
\end{itemize}

\begin{acknowledgements}
D. P.  was supported by the Centre National d'Etudes Spatiales (CNES).
Part of this work was financially supported by Grant-in-Aid for
Scientific Research (nos. 22700785 and 22340040) of Japan Society for
the Promotion of Science (JSPS). The authors would like to thank
T. M. Dame for making the unpublished CO data available to them to
perform this work.
\end{acknowledgements}

\begin{appendix}
\section{WCS to \Healpix ancillary data transformation}\label{sec:drizzeling}

The data used in this paper were transformed from their native
WCS (world coordinate system) local projection into the \Healpix
(Hierarchical Equal Area isoLatitude Pixelization) all-sky
pixelization using the method described here in this Appendix. This method
has also been used to produce \Healpix maps of a larger set of ancillary
data, particularly in the context of the
analysis of the Planck data \footnote{The ancillary data can
be accessed on \url {http://www.cesr.fr/~bernard/Ancillary/}}. We note
that the CO map used in this paper is not yet freely available, since the
corresponding data have not been made public.

The \Healpix format allows us to store ancillary data on a single grid
scheme on the sphere. This is advantageous since different data can
then be compared on a pixel by pixel basis, without the need to
project the data to a common grid. Ancillary data are available in the
WCS convention and are usually stored as Flexible Image
Transport System (\Fits) files. In the following, we refer to the
ancillary data as \Fits maps. It is usually impractical to return to the raw ancillary data
and directly reprocess the maps into the \Healpix pixelization.  It is
therefore necessary to project the \Fits format data onto the \Healpix
pixels. As the \Fits and \Healpix pixels do not match each other in position, size, and shape, this requires some kind of interpolation, which must be
done with the minimal loss of information and without altering the
photometry of the original ancillary data. Here, we use a mosaicking
method where we compute explicitly the surface of the pixel
intersections, and use these values as weights to construct the
\Healpix map.

\begin{figure}
  \centering
  \includegraphics[width=8cm]{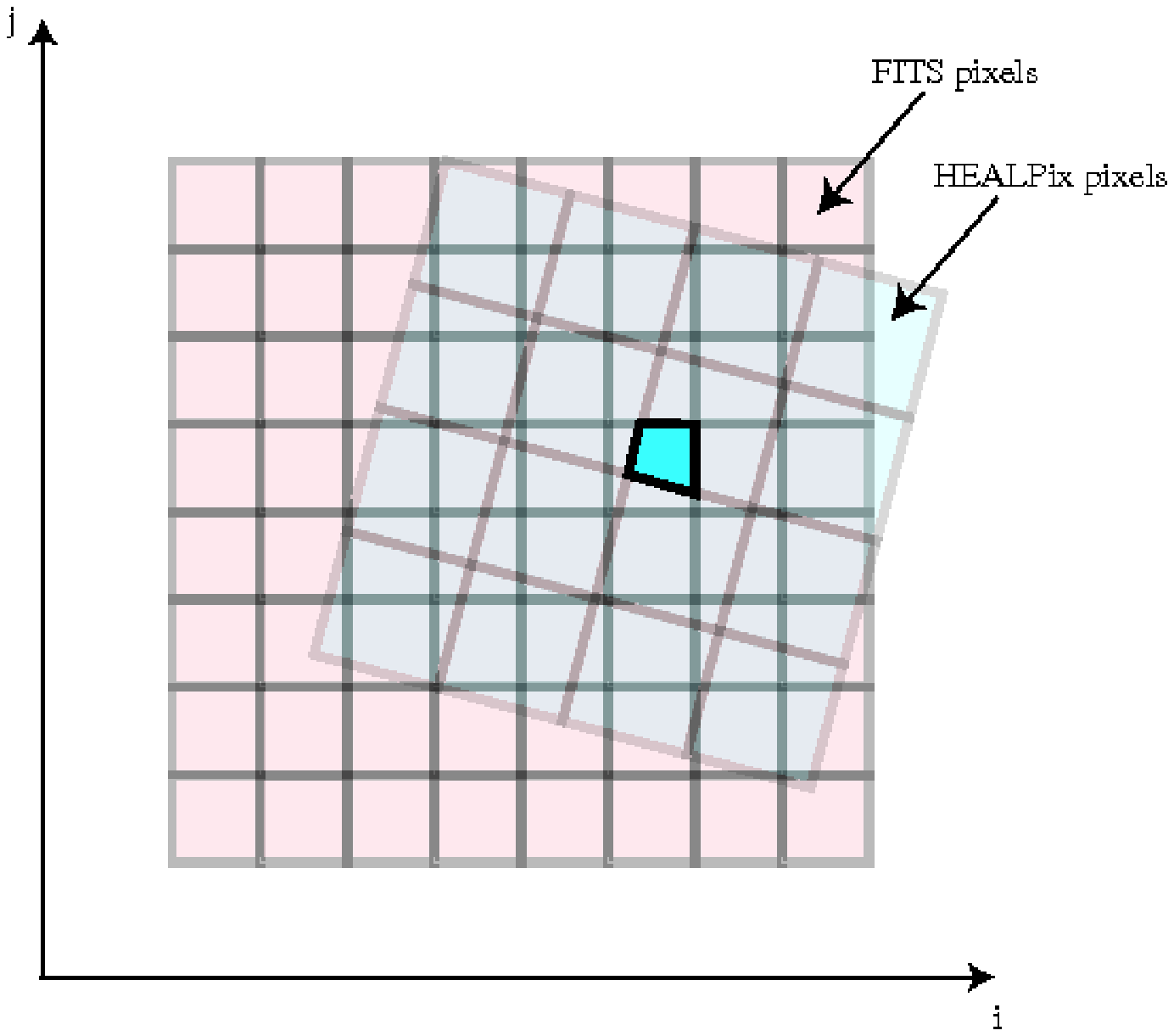}
    \caption{\label{fig:drizzle} Geometry of the \Healpix and \Fits pixels.}
\end{figure}

The \Healpix pixels projected onto a local \Fits map are shown in
Figure \ref{fig:drizzle}. The calculations are performed for a
\Healpix pixel size (given by the $\rm N_{side}$ \Healpix parameter) so
that the corresponding \Healpix pixel size matches the Shannon
criterion for the angular resolution of the data considered. This
ensures that no spatial information is lost in the conversion and that
the corresponding \Healpix and the \Fits pixels have similar sizes.  For a given \Fits
projection (i.e. from the astrometry information contained in the
\Fits header), we first identify the \Healpix pixels that intercept the
sky area covered by the \Fits map. For each of those \Healpix pixels,
we identify the \Fits pixels with a non-zero intersection.  We then
compute the surface fraction of the \Fits pixels intersecting the
\Healpix pixel $\rm S_{ih}^{if}$, where ``$\rm if$'' denotes the \Fits pixel
number and ``$\rm ih$'' denotes the \Healpix pixel number.

The calculation is performed in the pixel space of the \Fits map shown
in Figure \ref{fig:drizzle}. We first compute the sky coordinates of
the four corners of the considered \Healpix pixel.  We then transform
these coordinates into 2D pixel numbers (i,j) of the \Fits image using
standard routines and the astrometry information contained in the
\Fits header of the considered image. We then assume that the frontier
of the \Healpix pixel is a straight line in the \Fits image pixel
frame. Although this is not exactly true, this is a very small
approximation when the size of the \Healpix pixel is smaller or on the
order of the \Fits pixel size, which is always the case here. We
then compute the surface of the corresponding intersection polygon,
which has been normalized to
that of the \Healpix pixel surface $\rm S_{ih}^{if}$ \footnote{This calculation is
performed using the {\it polyfillaa} routine (see \url
{http://tir.astro.utoledo.edu/jdsmith/code/idl.php}) optimized for
fast clipping of polygons against a pixel grid.}.

The intersection geometry informations ($\rm if$ and $\rm S_{ih}^{if}$) are stored in a table
containing one entry per \Healpix pixel that has non-zero intersection with
the considered \Fits map. Calculations are carried and stored
separately for each \Fits map. We note that the above table can be
easily inverted to a table with one entry per \Fits pixel, giving
the \Healpix pixel numbers intercepting it ($\rm ih$) and the
corresponding surface fraction $\rm S_{if}^{ih}$, which in turn can be
used to perform reverse calculations projecting \Healpix maps onto local WCS
projection.

The above intersection fractions are then used to compute \Healpix ancillary data value
$\rm d_{ih}$, given the \Fits data $\rm d_{if}$ as
\begin{equation}
d_{ih}=\displaystyle\sum_{if} S_{if}^{ih} \times d_{if},
\end{equation}
where the summation is carried over all \Fits pixels intersecting
a \Healpix pixel $\rm ih$.

When the data is composed of a collection of individual \Fits maps, the calculation
are performed for each map separately, while maintaining a map of the total weight,
which allows us to evaluate average values in sky regions where individual maps overlap.
We note that virtually all WCS projection types can be used, including the
sixcube projection used for the COBE and FIRAS data.

\end{appendix}

\end{document}